\documentclass[sigconf]{acmart}
\AtBeginDocument{%
  }

\setcopyright{acmcopyright}
\copyrightyear{2018}
\acmYear{2018}
\acmDOI{XXXXXXX.XXXXXXX}

\acmConference[Conference acronym 'XX]{Make sure to enter the correct
  conference title from your rights confirmation emai}{June 03--05,
  2018}{Woodstock, NY}
\acmPrice{15.00}
\acmISBN{978-1-4503-XXXX-X/18/06}

\usepackage{booktabs}
\usepackage{multirow}
\usepackage{url}
\usepackage{enumitem}   
\usepackage{float} 

\definecolor{myred}{rgb}{0.68627451, 0.14117647, 0.09803922}

\newcommand{\myeq}[1]{\hyperref[eq:#1]{Eq.~(\ref*{eq:#1})}}
\newcommand{\mysec}[1]{\hyperref[sec:#1]{Section~\ref*{sec:#1}}}
\newcommand{\mytable}[1]{\hyperref[tab:#1]{Table~\ref*{tab:#1}}}
\newcommand{\myfig}[1]{\hyperref[fig:#1]{Fig.~\ref*{fig:#1}}}
\newcommand{\myappendix}[1]{\hyperref[appendix:#1]{Appendix}}
\newcommand{\myalg}[1]{\hyperref[alg:#1]{Algorithm~\ref*{alg:#1}}}
\newcommand{\myhyp}[1]{\hyperref[alg:#1]{\textsc{Hypothesis}~\ref*{hyp:#1}}}

\acmSubmissionID{(Resource) 5018}

\begin{document}

\title{KuaiRand: An Unbiased Sequential Recommendation Dataset with Randomly Exposed Videos}



\author{Chongming Gao}
\authornote{Three authors contributed equally to this research.}
\email{chongming.gao@gmail.com}
\orcid{0000-0002-5187-9196}
\affiliation{%
  \institution{University of Science and Technology of China}
  \country{}
}

\author{Shijun Li}
\authornotemark[1]
\email{lishijun@mail.ustc.edu.cn}
\affiliation{%
  \institution{University of Science and Technology of China}
  \country{}
}

\author{Yuan Zhang}
\authornotemark[1]
\email{yuanz.pku@gmail.com}
\affiliation{%
  \institution{Kuaishou Technology Co., Ltd.}
  \country{China}
}

\author{Jiawei Chen}
\email{sleepyhunt@zju.edu.cn}
\orcid{0000-0002-4752-2629}
\affiliation{%
  \institution{Zhejiang University}
  \country{}
}

\author{Biao Li}
\email{libiao@kuaishou.com}
\affiliation{%
  \institution{Kuaishou Technology Co., Ltd.}
  \country{}
}

\author{Wenqiang Lei}
\email{wenqianglei@gmail.com}
\affiliation{%
  \institution{Sichuan University}
  \country{}
}

\author{Peng Jiang}
\email{jiangpeng@kuaishou.com}
\affiliation{%
  \institution{Kuaishou Technology Co., Ltd.}
  \country{China}
}

\author{Xiangnan He}
\email{xiangnanhe@gmail.com}
\orcid{0000-0001-8472-7992}
\affiliation{%
  \institution{University of Science and Technology of China}
  \country{}
}

\renewcommand{\shortauthors}{Gao et al.}

\begin{abstract}
Recommender systems deployed in real-world applications can have inherent exposure bias, which leads to the biased logged data plaguing the researchers. A fundamental way to address this thorny problem is to collect users' interactions on randomly expose items, i.e., the missing-at-random data.
A few works have asked certain users to rate or select randomly recommended items, e.g., Yahoo! \cite{Marlin2009Collaborative}, Coat \cite{rs_treatment}, and OpenBandit \cite{saito2021open}. However, these datasets are either too small in size or lack key information, such as unique user ID or the features of users/items.
In this work, we present \emph{KuaiRand}, an unbiased sequential recommendation dataset containing millions of intervened interactions on randomly exposed videos, collected from the video-sharing mobile App, Kuaishou. Different from existing datasets, KuaiRand records 12 kinds of user feedback signals (e.g., click, like, and view time) on randomly exposed videos inserted in the recommendation feeds in two weeks. 
To facilitate model learning, we further collect rich features of users and items as well as users' behavior history. 
By releasing this dataset, we enable the research of advanced debiasing large-scale recommendation scenarios for the first time. Also, with its distinctive features, KuaiRand can support various other research directions such as interactive recommendation, long sequential behavior modeling, and multi-task learning. 
The dataset and its news will be available at \textcolor{magenta}{\url{https://kuairand.com}}.
\end{abstract}

\copyrightyear{2022}
\acmYear{2022}
\setcopyright{acmcopyright}
\acmConference[CIKM '22] {Proceedings of the 31st ACM International Conference on Information and Knowledge Management}{October 17--21, 2022}{Atlanta, GA, USA.}
\acmBooktitle{Proceedings of the 31st ACM International Conference on Information and Knowledge Management (CIKM '22), October 17--21, 2022, Atlanta, GA, USA}
\acmPrice{15.00}
\acmISBN{978-1-4503-9236-5/22/10}
\acmDOI{10.1145/3511808.3557624}



\maketitle
\section{Introduction}

\begin{figure}[!t]
  \tabcolsep=0pt
  \centering
  \includegraphics[width=1\linewidth]{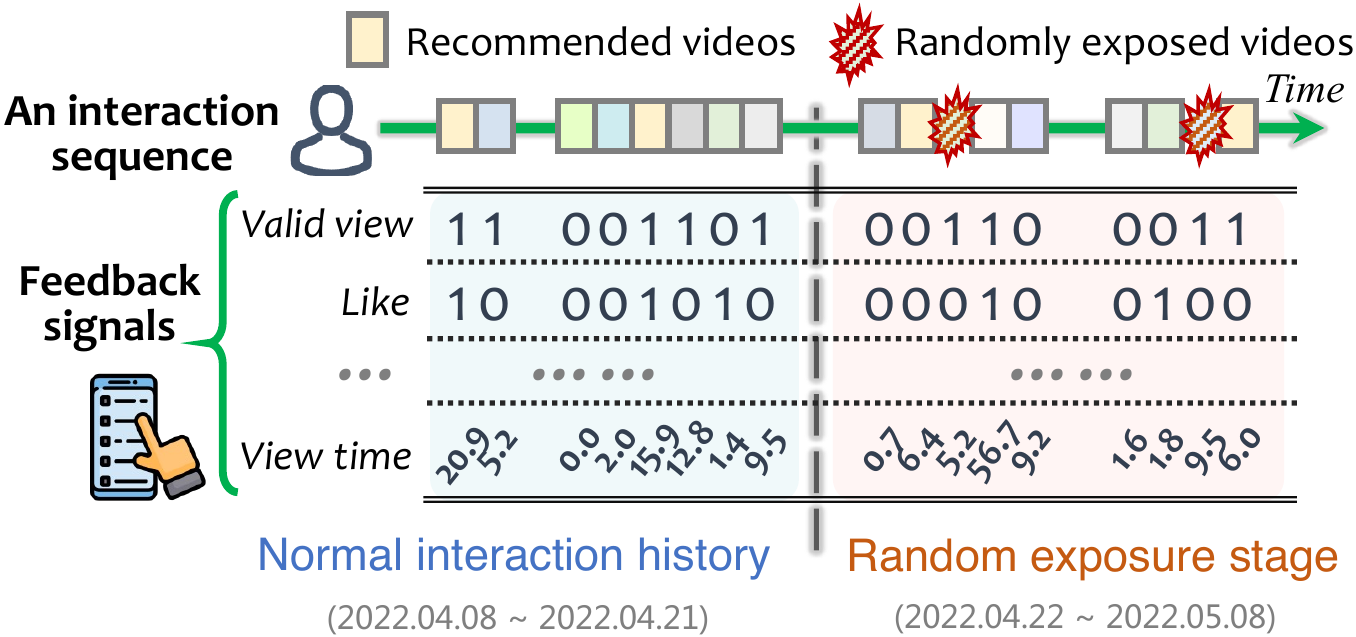}
   \vspace{-3mm}
  \caption{Illustration of the proposed \textcolor{myred}{\emph{KuaiRand}} dataset.}
   \vspace{-3mm}
  \label{fig:KuaiRand}
\end{figure}

\begin{table*}[!t]
\tabcolsep=4pt
\renewcommand\arraystretch{1.1}
\caption{Comparison of currently available datasets with randomly exposed data.}
\label{tab:datasets}
\begin{tabular}{ccrrrrrcc}
\hline
\textbf{Dataset}                                                  & \textbf{Collection policy}  & \multicolumn{1}{c}{\textbf{\# Users}} & \multicolumn{1}{c}{\textbf{\# Items}} & \multicolumn{1}{c}{\textbf{\# Interactions}} & \multicolumn{1}{c}{\textbf{\begin{tabular}[c]{@{}c@{}}\# User \\ features\end{tabular}}} & \multicolumn{1}{c}{\textbf{\begin{tabular}[c]{@{}c@{}}\# Item \\ features\end{tabular}}} & \textbf{Timestamp} & \textbf{Feedback}                                                                                         \\ \hline
\multirow{2}{*}{\textbf{Yahoo!R3 \cite{Marlin2009Collaborative}}} & \textbf{Normal interaction} & 15,400                                & 1,000                                 & 311,704                                      & 0                                                                                        & 0                                                                                        & No                 & Rating: [1,5]                                                                                             \\
                                                                  & \textbf{Random policy}      & 5,400                                 & 1,000                                 & 54,000                                       & 7                                                                                        & 0                                                                                        & No                 & Rating: [1,5]                                                                                             \\ \hline
\textbf{Yahoo!R6A \cite{li2010contextual}}                        & \textbf{Random policy}      & No ID                                 & 271                                   & 45,811,883                                   & 6                                                                                        & 6                                                                                        & Yes                & Click                                                                                                     \\
\textbf{Yahoo!R6B \cite{li2010contextual}}                        & \textbf{Random policy}      & No ID                                 & 652                                   & 27,777,695                                   & 136                                                                                      & 0                                                                                        & Yes                & Click                                                                                                     \\ \hline
\multirow{2}{*}{\textbf{Coat \cite{rs_treatment}}}                & \textbf{User self-selected} & 290                                   & 300                                   & 6,960                                        & 4                                                                                        & 4                                                                                        & No                 & Rating: [1,5]                                                                                             \\
                                                                  & \textbf{Random policy}      & 290                                   & 300                                   & 4,640                                        & 4                                                                                        & 4                                                                                        & No                 & Rating: [1,5]                                                                                             \\ \hline
\multirow{2}{*}{\textbf{Open Bandit \cite{saito2021open}}}        & \textbf{Bernoulli TS}       & No ID                                 & 80                                    & 24,011,308                                   & 4                                                                                        & 4                                                                                        & Yes                & Click                                                                                                     \\
                                                                  & \textbf{Random policy}      & No ID                                 & 80                                    & 2,691,861                                    & 4                                                                                        & 4                                                                                        & Yes                & Click                                                                                                     \\ \hline
\multirow{2}{*}{\textbf{\textcolor{myred}{KuaiRand-27K}}}          & \textbf{15 policies}        & 27,285                                   &  32,038,725                                     & 322,278,385                                             &   30                                                                                       & 62                                                                                         & Yes                & \multirow{6}{*}{\begin{tabular}[c]{@{}c@{}}12 signals\\ (e.g., click, \\ like, hate,\\ view time)\end{tabular}} \\
                                                                  & \textbf{Random policy}      & 27,285                                   & 7,583                                      & 1,186,059                                             &         30                                                                                 &  62                                                                                    & Yes                &                                                                                                           \\
\multirow{2}{*}{\textbf{\textcolor{myred}{KuaiRand-1K}}}           & \textbf{15 policies}        & 1,000                                    & 4,369,953                                      &  11,713,045                                            &  30                                                                                       &  62                                                                                     & Yes                &                                                                                                           \\
                                                                  & \textbf{Random policy}      & 1,000                                    & 7,388                                      & 43,028                                             &      30                                                                                  & 62                                                                                         & Yes                &                                                                                                           \\
\multirow{2}{*}{\textbf{\textcolor{myred}{KuaiRand-Pure}}}           & \textbf{15 policies}        & 27,285                                    & 7,551                                      &  1,436,609                                            &  30                                                                                       &  62                                                                                     & Yes                &                                                                                                           \\
                                                                  & \textbf{Random policy}      & 27,285                                    & 7,583                                      & 1,186,059                                             &      30                                                                                  & 62                                                                                         & Yes                &                                                                                                           \\ \hline
\end{tabular}
\end{table*}

Recommender systems have become a powerful tool for tech companies to connect users with appropriate items or services. 
Although developing and evaluating a recommendation model with real users are effective, online A/B testing is time- and money-consuming and entails the risk of hurting customer engagement, thus impairing the platform's value proposition \cite{schnabel2018short,gilotte2018offline}. Hence, most of the efforts have been put into developing advanced recommendation models on the offline data generated from the interaction logs. 

However, the bottleneck becomes the offline data \emph{per se}. Since the highly sparse data comes from the collected historical interactions between users and the system, it is unavoidably affected by the exposure bias \cite{chen2020bias,10.1145/3041021.3054202,10.1145/3397271.3401083}. The biased data has catastrophic effects on researchers and practitioners of recommender systems. It leads to inconsistent online and offline results of the model \cite{rs_treatment} and gives rise to filter bubbles \cite{gao2022cirs}, which can incur heavy losses for the companies and hinder the development of recommender systems.


\subsection{Open Problem: Collecting Random Data}
The fundamental problem lies in the fact that we have no knowledge about the massive missing interactions in the offline data \cite{gao2022kuairec}. To tackle the problem at the root, we have to collect the user preferences regarding those previously NOT exposed items. \citet{gao2022kuairec} collected a fully-observed dataset with each user's preferences on each item known. 
However, this dataset can inevitably be biased towards popular items due to its extremely high density.

A more practical way to obtain truly unbiased data is to collect users' feedback on uniformly sampled items, i.e., the missing-at-random (MAR) data \cite{Yang_2018_unbiased_evaluation}. It has been demonstrated that we can conduct an unbiased offline evaluation on the MAR data \cite{Yang_2018_unbiased_evaluation,saito_unbiased}.


Few works have considered this solution and collected user feedback on randomly exposed items. The well-known datasets Yahoo! \cite{Marlin2009Collaborative} and Coat \cite{rs_treatment} asked a set of people to explicitly rate a certain number of randomly selected items in the score range from $1$ to $5$. However, the scales of the two datasets are quite limited --- only $54,000$ ratings in Yahoo! and $4,640$ ratings in Coat for the randomly selected items. Besides, both of them lack some key information such as user/item features and timestamp. The recently introduced dataset OpenBandit \cite{saito2021open} collected users' clicks on items from a slate of three items. Each slate is generated by a multi-armed bandit policy or a random policy. However, it only evolves $80$ items, and the user identification (ID) information is removed. These insufficient information problems make existing datasets fail to satisfy the requirement of many recommendation tasks.

\subsection{Contributions: The KuaiRand Dataset}
Our goal is to enable unbiased offline evaluation in real-world recommendation scenarios. To this end, we present \emph{KuaiRand}, a large-scale sequential recommendation dataset, collected from Kuaishou App\footnote{\url{https://www.kuaishou.com/cn}}, one of the China's largest video-sharing Apps with more than $300$ million daily active users. 
Different from previous datasets, it intervenes in the recommendation policies by randomly inserting randomly chosen videos in the normal recommendation feed for two weeks without user awareness. Hence, we can collect genuine unbiased user reflections on these randomly exposed videos. We recorded all $12$ kinds of feedback signals, such as \emph{click}, \emph{adding to favorites}, and \emph{view time}. To facilitate model learning, we further collected users' historical behaviors as well as rich side information of users/items. The process is illustrated in \myfig{KuaiRand} and the detailed information can be referred to \mytable{datasets}. 

It is worth mentioning that KuaiRand is the first unbiased sequential recommendation data where we intervene in the original recommendation policy millions of times by inserting random items. 
The unbiased sequential data enables us to carry out the unbiased offline evaluation \cite{Yang_2018_unbiased_evaluation,saito_unbiased}, thus facilitating the research of debiasing \cite{chen2020bias} in large-scale recommendation scenarios. 
Furthermore, with its distinctive features, KuaiRand can naturally facilitate a variety of other recommendation tasks, such as interactive recommendation \cite{xinxin20,gao2022cirs}, long sequential behavior modeling \cite{qipikdd,qipicikm}, and multi-task learning \cite{cai2022constrained,yang2022cross}. 
\section{Related Work}
\label{sec:related}

\subsection{Offline Evaluation in Recommendation} 

Online A/B tests are effective in tech companies to assess the performances of recommender models \cite{gilotte2018offline}. However, it usually consumes much time and money, which makes it impractical for academic researchers to conduct the evaluation online.

Therefore, researchers usually resort to offline evaluation, which includes:
(1) computing traditional indicators (such as Precision, Recall, and NDCG \cite{NDCG}) on the test set;
(2) conducting \emph{off-policy evaluation} or \emph{counterfactual reasoning} \cite{bottou2013counterfactual,swaminathan2015counterfactual} to perform an offline A/B test \cite{gilotte2018offline};
(3) using user simulation techniques \cite{simulation2020kdd,ie2019recsim} to evaluate the recommender in the online A/B test.

However, computing traditional indicators cannot answer how the recommender performs on the massive missing data, i.e., the user-item pairs which have not occurred in the test set. Though off-policy evaluation and user simulation can solve this pitfall, they suffer the problem of high-variance issue \cite{saito2021open} and additionally introduced error \cite{gao2021advances}, respectively.

\subsection{Datasets with Randomly Exposed Items}
A way to fundamentally solve this problem in offline evaluation is to collect unbiased data, i.e., to elicit user preferences on the randomly exposed items.
We briefly introduce classic existing high-quality datasets that contain unbiased randomly-sampled data used for offline evaluation in recommendation.
\begin{itemize}
    \item \textbf{Yahoo!R3} \cite{Marlin2009Collaborative}. It contains the conventional missing-not-at-random (MNAR) data from $15,400$ users on $1,000$ items in total. It also contains a set of missing-complete-at-random (MCAR) data by asking $5,400$ users to give ratings on 10 items that are randomly selected from the 1,000 items. However, it does not contain timestamps or the features of users and items. 
    \item \textbf{Yahoo!R6} \cite{li2010contextual}. It has two versions, namely R6A and R6B, both of which have only the random policy to expose hundreds of news millions of times. Besides, all user IDs are removed, which forbids it to perform basic collaborative filtering tasks in recommendation.
    \item \textbf{Coat} \cite{rs_treatment}. It collects the ratings of $290$ users on $24$ self-selected items and $16$ randomly-selected items from a total of $300$ items. Similar to Yahoo!R3, the scale of it is too small compared with other datasets in real-world recommendation scenarios. Neither does it has the timestamp to support research on sequential recommendation.
    \item \textbf{Open Bandit Dataset} \cite{saito2021open}. It contains interactions collected from two logged bandit policies: a Bernoulli Thompson Sampling policy and a uniform (random) policy. There are approximately $26$ million interactions collected by users' clicks on 80 items. The disadvantage of this data is the lack of user ID, making it hard to support the research on sequential recommendation. In addition, the fact that it has only 80 items limits its reliability in evaluating most recommendation algorithms.
\end{itemize}
The statistics of these datasets are summarized in \mytable{datasets}. Compared with them, our proposed KuaiRand has the following key advantages: 
(1) it has the most comprehensive side information including explicit user IDs, interaction timestamps, and rich features for users and items;
(2) it has 15 policies with each catered for a special recommendation scenario in Kuaishou App;
(3) we introduce 12 feedback signals for each interaction to describe the user's comprehensive feedback;
(4) each user has thousands of historical interactions on average.
These distinctive characteristics can enable a lot of potential research directions.

\section{Data Description}
\label{sec:data}

Here, we introduce the basic characteristics of Kuaishou App and the process of how we collect the KuaiRand dataset.

\subsection{Characteristics of Kuaishou App}
Kuaishou App is one of the largest short video-sharing platforms in China. It has amassed over $300$ million daily-active users with an average spent time of nearly $128.1$ minutes per day \cite{kuaishou2022q1}. Different from other recommendation scenarios such as e-commerce, users in short video recommendation platforms can generate very long interaction sequences every day. There are also many new videos published every day, hence the videos become outdated very quickly. 

\begin{figure}[!t]
  \tabcolsep=0pt
  \centering
  \includegraphics[width=1\linewidth]{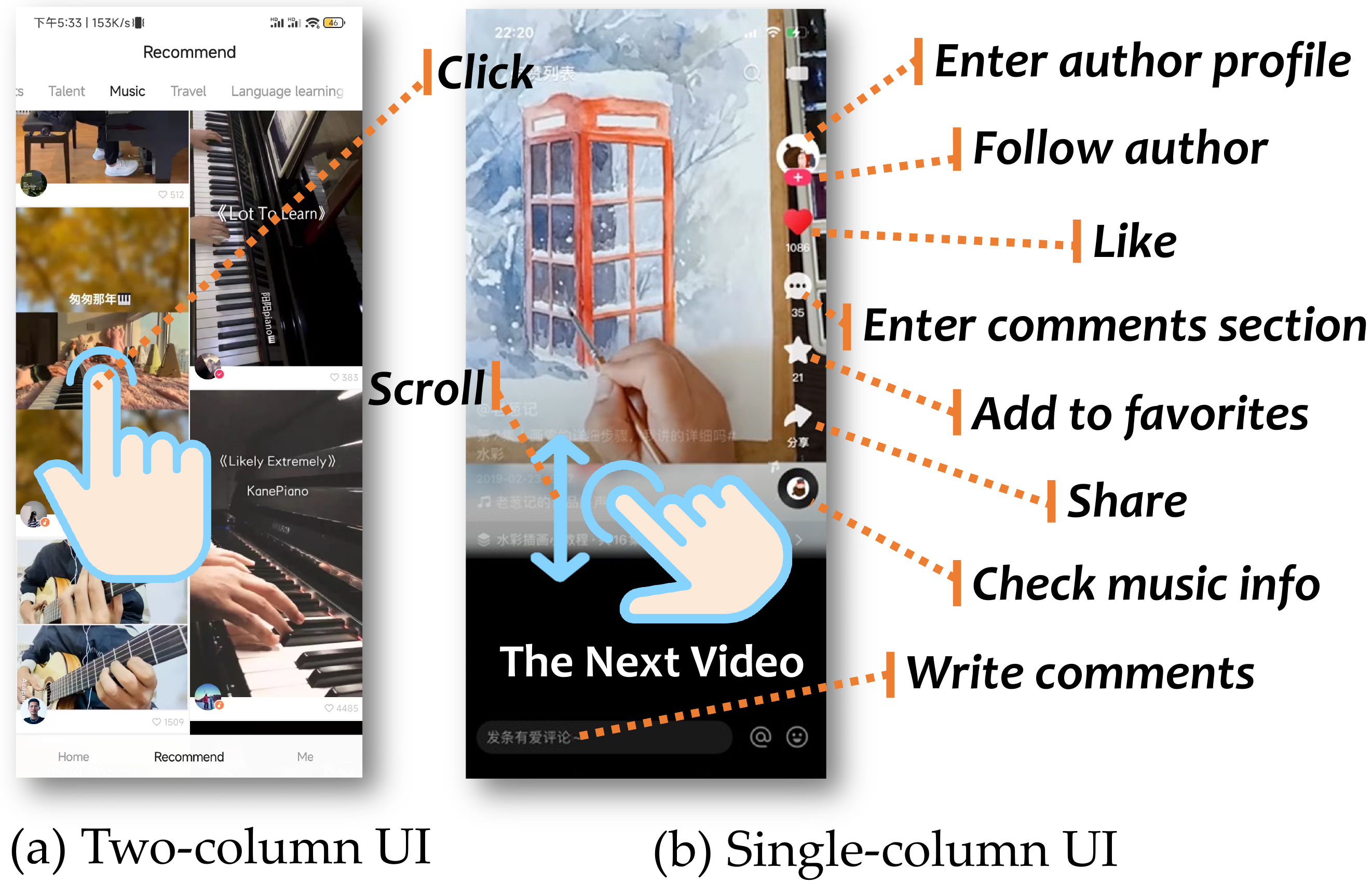}
   \vspace{-3mm}
  \caption{Illustration of two recommendation scenarios in Kuaishou App. A variety of user behaviors are marked.}
   \vspace{-3mm}
  \label{fig:ui}
\end{figure}

The videos are mainly recommended to users in different 15 scenarios, e.g., under different user interfaces (UIs) in Kuaishou App or the embedded UI in other Apps such as WeChat Mini Program. Each scenario has distinctive rules and objectives. Therefore, each scenario has its own recommendation policy and there are 15 policies in total. We illustrate the most two commonly used UIs in two scenarios in \myfig{ui}.

On the left-hand side lies the main UI, where the thumbnails of recommended videos are organized in a two-column cascade style. In this UI, users can scroll up or down until they find the video they want to view, and then view this video by clicking the thumbnail. Therefore, \emph{click} is a critical signal indicating user interest in this UI style. On the right-hand side, we can see the single-column recommendation cascading UI. It is the most commonly used UI. The video will automatically play whenever switching to this UI. There is no click signal in the two-column UI, but users can choose to stop viewing this video by exiting this UI or scrolling to the last/next video. Therefore, \emph{view time} in this UI becomes an important signal to reflect user satisfaction. Besides, there are other signals representing users' behaviors when they interact with this video, including \emph{like}, \emph{enter author profile}, \emph{follow author}. These feedback signals correspond to the buttons on this UI (\myfig{ui}), so we can easily collect these timely signals.

\subsection{Data Collection}
Our goal is to collect the sequential recommendation data with randomly exposed videos inserted. We achieve this goal via the following three steps:
\begin{enumerate}[label=(\roman*)]
  \item We sample a batch of videos and filter out the spam such as advertisements. There are $7,583$ items in total. For the target users, we randomly select a batch of users and remove robots, which includes over $200,000$ real users. 
  \item Each time when the recommender system recommends a video list to a user, we decide whether to insert a random item with a fixed probability\footnote{We hide the exact value for a commercial security reason, i.e., to avoid leaking out the key parameters of the online recommendation strategy.}. If the answer is yes, then we intervene in the recommendation list by randomly selecting one video from this list and replacing it with a random item uniformly sampled from the $7,583$ items.
  This process lasts two weeks (from April 22, 2022, to May 08, 2022). We record $12$ feedback signals reflecting the comprehensive behaviors of these users. We remove the users that have been exposed to less than $10$ randomly exposed videos for faithful evaluation. There are $27,285$ users retained. All $7,583$ items have been inserted at least once, and the total number of random interventions is $1,186,059$.
  \item To facilitate model learning on this dataset, we further retrieve the historical user interaction of these $27,285$ users (without intervention by the random policy), which also lasts two weeks (from April 08, 2022, to April 21, 2022). There are $32,038,725$ items and $322,278,385$ times of normal recommendations included in the one-month dataset.
  In addition, we further collect the side information including the key features of these users and items.
\end{enumerate}
So far, we have collected the KuaiRand dataset.

\subsection{Statistics and Usage}
The basic statistics of KuaiRand are summarized in \mytable{datasets}. We will release three versions (\emph{KuaiRand-27K}, \emph{KuaiRand-1K}, and \emph{KuaiRand-Pure}) for different usages, which are described in \myappendix{versions}.

\begin{figure}[!t]
  \tabcolsep=0pt
  \centering
  \includegraphics[width=1\linewidth]{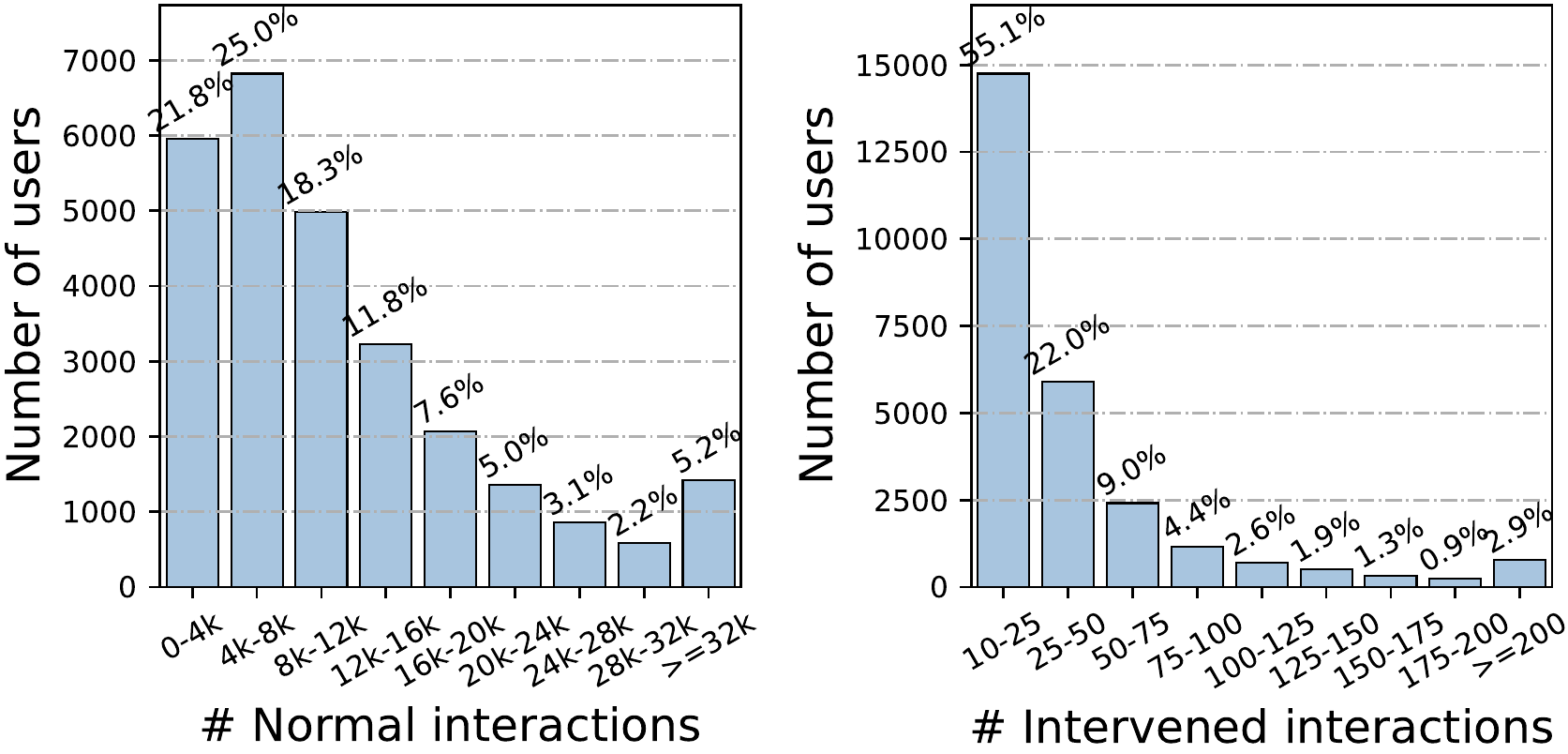}
  \caption{Distribution of users w.r.t. the number of normal interactions (i.e., receiving ordinary recommendations) and intervened interactions (i.e., exposed to random videos).}
   \vspace{-3mm}
  \label{fig:distribution}
\end{figure}

For \emph{KuaiRand-27K}, we visualize the distribution of the users w.r.t. the numbers of interactions \myfig{distribution}. An interaction can either be a normally recommended video or a randomly chosen video. From the figure, we can see that most users in this dataset have viewed thousands of recommended videos in one month. Each user has more than 10 times viewed random videos. 
For the detailed usage and news of this dataset, please refer to \url{https://kuairand.com}.

\section{Potential Research Directions}
\label{seq:future}

Through releasing KuaiRand, we offer the opportunity to carry out the debiasing task in the large-scale sequential recommendation for the first time. 
\smallskip
\begin{itemize}[nosep,leftmargin=*]
	\item \textbf{Debiasing in Recommendation}. Recommender systems suffer from various biases in the data collection stage \cite{chen2020bias}. Most existing datasets are very sparse and affected by user-selection bias \cite{Marlin2009Collaborative} or exposure bias (previous model bias) \cite{10.1145/3397271.3401083}. It is of critical importance to develop models that can alleviate biases. To evaluate the models, we need reliable unbiased data. KuaiRand is the first dataset that inserts the random items into the normal recommendation feeds with rich side information and all item/user IDs provided. With this authentic unbiased data, we can evaluate and thus improve the recommender policy.
\end{itemize}

\smallskip
With its distinctive features, KuaiRand can further support the following promising research directions in recommendation.
\smallskip
\begin{itemize}[nosep,leftmargin=*]
    \item \textbf{Off-policy Evaluation (OPE).} OPE is proposed to overcome the difficulties in conducting online A/B testing in recommendation. OPE aims to estimate the performance of a counterfactual policy using only log data generated by past policies \cite{saito2021open}. It requires that we have the log sequences rather than merely positive samples. KuaiRand satisfies this requirement. It recorded logs under 15 policies catered for different situations. Therefore, we can leverage OPE and conduct offline A/B tests to improve our recommendation policies \cite{gilotte2018offline}.
	\item \textbf{Interactive Recommendation.} An interactive recommender system (IRS) is usually formulated as a decision-making process to pursue long-term success in recommendation. IRS built on reinforcement learning has shown superiority to the traditional supervised frameworks \cite{wang2022best,xinxin20,gao2022cirs}. KuaiRand contains the sequential logs generated from different policies that can naturally support the research on IRSs with the help of OPE and user simulation techniques \cite{gao2021advances,tutorial_CRS}.
	\item \textbf{Long Sequential Behavior Modeling.} In practice, traditional sequential recommendation models, such as DIN \cite{zhou2018deep} and DIEN \cite{zhou2019dien}, can only model a maximum of hundreds of items in a sequence. Recently, some works have demonstrated that modeling users' long sequential behaviors is very effective in lifting click-through rate (CTR) \cite{qipikdd,qipicikm}. KuaiRand has thousands of interactions for each user, which is perfect for the long sequential user modeling task.
	\item \textbf{Multi-Task Learning.} In real-world recommender systems, user satisfaction can be reflected in multiple behaviors, e.g., clicks, writing down positive comments, and staying longer. Usually, these metrics cannot be pursued at the same time in one model. Therefore, multi-task learning or multi-object learning has received growing attention \cite{mtl1,cai2022constrained,yang2022cross}. KuaiRand recorded 12 feedback signals for each interaction, which promotes the research on learning with more than ten tasks. In contrast, most existing datasets in multi-task recommendation \cite{lin2019pareto, ma2018entire} only provide less than 5 signals.
\end{itemize}

\section{Acknowledgements}
This work is supported by the National Natural Science Foundation of China (U19A2079, 62102382).

\bibliographystyle{ACM-Reference-Format}
\bibliography{KuaiRand}

\clearpage
\appendix

\begin{figure*}[H]
  \tabcolsep=0pt
  \centering
  \twocolumn[{\includegraphics[width=.95\linewidth]{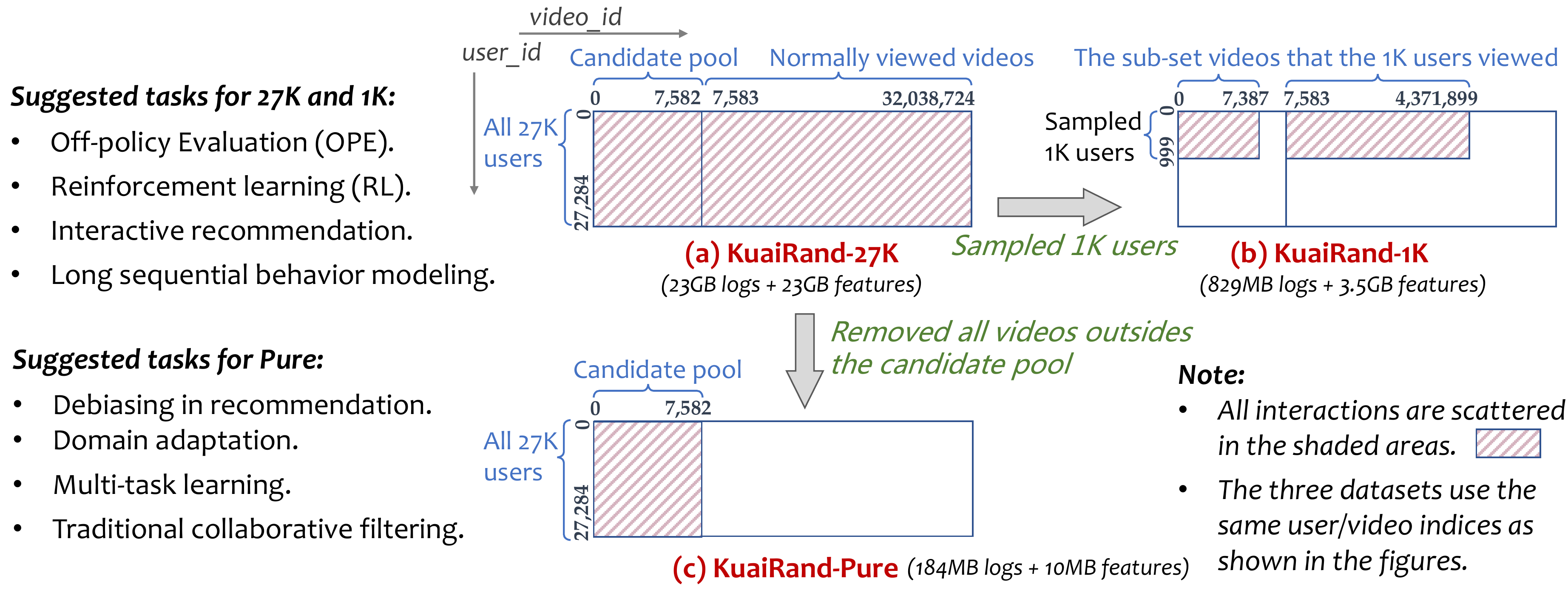}   
  \caption{Illustration of user/video ID spaces of the three versions of \emph{KuaiRand} and the suggested tasks.}
   \vspace{5mm}
  \label{fig:three}
  }]
\end{figure*}


\section{Versions and Suggestions}
\label{appendix:versions}

Considering its scale is intimidating for some researchers, we release three versions of the data for different usages. 
We give a brief comparison and suggestions for which version to use as follows:
\begin{itemize}
  \item \textbf{KuaiRand-27K (23GB logs + 23GB features)}: The complete KuaiRand data contains over 27K users and 32 million videos. If your research needs rigorous sequential logs, such as off-policy evaluation (OPE), Reinforcement learning (RL), or long sequential recommendation.
  \item \textbf{KuaiRand-1K (829MB logs + 3.5GB features)}: We uniformly select 1,000 users from KuaiRand-27K along with their logs and remove all irrelevant users/videos. There are about 4 million videos rest. If your computing resources are not enough to handle the whole KuaiRand-27K data, you can use this one. 
  \item \textbf{KuaiRand-Pure (184MB logs + 10MB features)}: We only keep the logs for the 7,582 videos in the candidate pool and remove all logs related to other videos. You can use this one if rigorous sequential information is not required in your research. For example, if you are studying debiasing in collaborative filtering models, domain adaptation, or multi-task modeling in recommendation.
\end{itemize}
The statistics of the three datasets are summarized in \mytable{datasets}. We visualize the user/video ID spaces of the three versions in \myfig{three}.

\end{document}